\documentclass[a4wide, 12pt]{article}

\usepackage{amsmath, amssymb, amsfonts}  
\usepackage[mathscr]{eucal}
\usepackage{cases}

\usepackage{bm}
\usepackage{bbm}
\usepackage{dsfont}
\usepackage{xcolor}
\usepackage{ifpdf}

\usepackage{graphics} 
\usepackage{graphicx}           
\graphicspath{{./Images/}}

\usepackage{enumitem}  
\setlist[description]{leftmargin=\parindent}

\usepackage{subfigure}

\usepackage{ifpdf}

\newcommand*\mcapinn[2]{\vcenter{\hbox{$\mathsurround=0pt
  \ifx\displaystyle#1\textstyle\else#1\fi\bigcap$}}}

\newcommand*\mcupinn[2]{\vcenter{\hbox{$\mathsurround=0pt
  \ifx\displaystyle#1\textstyle\else#1\fi\bigcup$}}}

\DeclareFontFamily{OT1}{pzc}{}
\DeclareFontShape{OT1}{pzc}{m}{it}{<-> s * [1.200] pzcmi7t}{}
\DeclareMathAlphabet{\mathpzc}{OT1}{pzc}{m}{it}

\usepackage{theorem}
\newtheorem{theorem}{Theorem}
\newtheorem{definition}{Definition}
\newtheorem{corollary}{Corollary}
\newtheorem{lemma}{Lemma}

\newtheorem{proposition}{Proposition}
\newtheorem{assumption}{Assumption}

\def\qed{\hfill \vrule height 5pt width 5pt depth 0pt \medskip}
\newcommand{\proof}{\noindent {\bf Proof. }}

{\theorembodyfont{\upshape}
	\newtheorem{remark}{Remark}
	\newtheorem{example}{Example}
	
}

\newcommand{\bfone}{\textbf{1}}

\newcommand{\beq}{\begin{equation}}
\newcommand{\eeq}{\end{equation}}
\newcommand{\beqa}{\begin{eqnarray}}
\newcommand{\eeqa}{\end{eqnarray}}
\newcommand{\beqan}{\begin{eqnarray*}}
\newcommand{\eeqan}{\end{eqnarray*}}

\newcommand{\pde}[2]{ \frac{\partial #1}{\partial #2} }

\newcommand{\bite}{\begin{itemize}}
\newcommand{\eite}{\end{itemize}}
\newcommand{\benu}{\begin{enumerate}}
\newcommand{\eenu}{\end{enumerate}}

\title{\LARGE \bf A dynamical approach to privacy preserving average consensus\thanks{Work supported in part by a grant from the Swedish Research Council (grant n. 2015-04390).}}

\author{Claudio Altafini\thanks{C. Altafini is  with the Division of Automatic Control, Department of Electrical Engineering, Link\"{o}ping University, SE-58183 Link\"{o}ping, Sweden. E-mail:   claudio.altafini@liu.se}
		}
\date{}

\begin{document}

\maketitle

\begin{abstract}
In this paper we propose a novel method for achieving average consensus in a continuous-time multiagent network while avoiding to disclose the initial states of the individual agents. 
In order to achieve privacy protection of the state variables, we introduce maps, called output masks, which alter the value of the states before transmitting them.
These output masks are local (i.e., implemented independently by each agent), deterministic, time-varying and converging asymptotically to the true state. 
The resulting masked system is also time-varying and has the original (unmasked) system as its limit system. 
It is shown in the paper that the masked system has the original average consensus value as a global attractor. However, in order to preserve privacy, it cannot share an equilibrium point with the unmasked system, meaning that in the masked system the global attractor cannot be also stable.
\end{abstract}



\section{Introduction}
Preserving privacy in a multiagent system means performing a computation in a distributed manner among the agents of a network without revealing the individual values that the agents contribute to the computation process. 
For instance a privacy preserving average consensus problem consists in computing the mean of the state variables of the agents without disclosing the initial state values to the other agents or to external observers. 
Several approaches have been proposed in recent years for this task. 
One of them relies on differential privacy \cite{Dwork:2006:DP:2097282.2097284,Dwork:2014:AFD:2693052.2693053}, which consists in adding to the state being transmitted by an agent a noise from an appropriate source. 
In this way, even if the value is publicly broadcasted, the knowledge that an observing agent can acquire of the true state is limited to a predetermined precision.
In the average consensus problem, this method has been investigated for instance in \cite{Cortes7798915,GUPTA20179515,Huang:2012:DPI:2381966.2381978,NOZARI2017221}.
Another approach relies on cryptography. 
Encrypted messages can be exchanged among the agents in various ways, e.g. through trusted third parties \cite{Lazzeretti6855039}, obfuscation \cite{Ambrosin:2017:OOB:3155100.3137573}, or through distributed cryptography schemes \cite{Liu2019Dynamical,RuanW17}. For instance in \cite{Manitara6669251,Mo7465717,Rezazadeh2018Privacy} the encryption is realized as a perturbation with zero sum (or integral) over time.
A third approach is based on understanding what is observable and what is not at a node \cite{Alaeddini7963642,Pequito7039593}, and on trying to guarantee privacy as a loss of observability.  

The scope of this paper is to propose a novel approach for privacy-preserving average consensus. 
Our approach is inspired by system-theoretical considerations and relies crucially on interpreting a distributed computation problem like consensus as a dynamical system, hence we refer to it as {\em dynamical privacy}.
We push the idea of ``lack of observability as a form of privacy'' to its extreme, by defining output maps that we call {\em output masks} which are altering (or ``masking'') the internal state of an agent before it is publicly transmitted. As these output masks remain private to each agent, they do not allow for any form of observability of the state of the agents. 
Our output masks are deterministic, time-varying transformations that each agent can implement independently, and that asymptotically converge to the true internal state.
The resulting masked system is also a time-varying system.
Its characteristic feature is that its limit system \cite{Artstein1976} corresponds to the original (unmasked) system, i.e., to the average consensus dynamics. 
If the average consensus problem being investigated is on a static graph, then the masked system is a case of asymptotically autonomous time-varying system \cite{Artstein1976,Markus1956}. 

We show in the paper that for suitably chosen output masks, in the consensus problem privacy protection can always be guaranteed even in the worst cases, for instance in the degenerate situation of all agents having the same initial state (and hence being already at the consensus value). 
In order to do so, an output mask must be able to escape neighborhoods of any point (i.e., it cannot be stable), feature which can be achieved by using output masks which are inhomogeneous in the state they are hiding.
This is reminiscent of the additive noise term used in differential privacy. 
A consequence of having inhomogeneous output maps is that the masked system lacks equilibria (again, fixed points in the masked dynamics would lead to breaching of privacy for certain initial conditions). 
In spite of the absence of equilibria, we shown in the paper that our time-varying masked system has the average consensus value as a global uniform attractor for its dynamics. 
In fact, the masked system converges asymptotically to the original consensus problem, and for all times a conservation law like preservation of the average of the true states remains valid also on the masked system. 

Technically, global attractivity can be shown using the same Lyapunov function of a standard consensus problem, but without requiring its derivative along the trajectories of the masked system to remain nonpositive. 
As already mentioned, the lack of stability around the average consensus value is needed to guarantee indiscernibility of the initial conditions from the masked outputs. 
Global attractivity is obtained if the derivative of the Lyapunov function is upper bounded by terms that decay to $0$ asymptotically. 

In this paper, the continuous-time version of the average consensus problem is considered. Unlike \cite{Rezazadeh2018Privacy}, we do not require that out perturbations have time integral that asymptotically tend to 0.
However, as in \cite{Rezazadeh2018Privacy}, we must impose that in-neighbors of the agents are not completely contained into each other.
The other conditions under which our our dynamical privacy guarantees privacy protection are mild and reasonable.
Namely, we require that only the internal state and the parameters of the output mask are kept private to each agent, while the masked output is communicated to the first neighbors on the interaction graph, and the Laplacian of the problem can be publicly available. 
It is also worth observing that breaching the privacy of an agent does not compromise that of the remaining agents, as each output mask is created locally by each agent.

\section{Preliminaries}
\label{sec:prelim}

A continuous function $ \alpha \, : \, [0, \infty) \to [ 0, \, \infty) $ is said to belong to class $ \mathcal{K}_\infty $ if it is strictly increasing and $ \alpha(0) =0$. 
Subclasses of $ \mathcal{K}_\infty $ which are homogeneous polynomials of order $ i$ will be denoted $ \mathcal{K}_\infty^i $: $ \alpha(r) = a r^i $ for some constant $ a>0$.
A continuous function $ \zeta \, : \, [0, \infty) \to [ 0, \, \infty) $ is said to belong to class $ \mathcal{L} $ if it is decreasing and $ \lim_{t\to \infty} \zeta(t) =0$. 
In particular, we are interested in $ \mathcal{L} $ functions that are exponentially decreasing: $ \zeta(t) = a e^{-\delta t} $ for some $ a>0$ and $ \delta>0$. We shall denote such subclass $ \mathcal{L}^e \subset  \mathcal{L} $.
A continuous function $ \beta \, : \, [0, \infty) \times [0, \, \infty) \to [ 0, \, \infty) $ is said to belong to class $ \mathcal{KL}_\infty^{i,e} $ if the mapping $ \beta(r, \, t ) $ belongs to class $ \mathcal{K}_\infty^i $ for each fixed $ t $ and to class $ \mathcal{L}^e $ for each fixed $ r$, i.e., $ \beta(r, \, t) = a r^i  e^{-\delta t }$ for some $ a>0$ and $\delta>0$.

Consider
\beq
\dot x = g(t, \, x ), \qquad x(t_o) =x_o 
\label{eq:ode_f(t,x)}
\eeq
where $ g\, : \, \mathbb{R}_+ \times \mathbb{R}^n \to \mathbb{R}^n$ is Lipschitz continuous in $ x$, measurable in $ t$, and such that for each $ x_o \in \mathbb{R}^n $ and each $ t_o \in \mathbb{R}_+ $ the solution of  \eqref{eq:ode_f(t,x)}, $ x(t,  \, x_o ) $, exists in $[0, \, \infty)$.  A point $ x^\ast \in \mathbb{R}^n $ is an equilibrium point of \eqref{eq:ode_f(t,x)} if $ g(t, \, x^\ast ) =0 $ for a.e. \footnote{almost every, i.e., except for at most a set of Lebesgue measure $0$.} $  t \geq t_o$. 

A point $ x^\ast \in \mathbb{R}^n $ is {\em uniformly globally attractive} for \eqref{eq:ode_f(t,x)} if for each $ \nu >0 $ there exists $ T = T(\nu) > 0 $ such that for each solution $ x(t, x_o ) $ of \eqref{eq:ode_f(t,x)} it holds that $ \| x(t, \, x_o ) - x^\ast \| < \nu$ for each $ t > t_o + T $, each $ x_o \in \mathbb{R}^n $ and each $ t_o \geq 0$. 
In particular, if $ x^\ast $ is a uniformly global attractor for \eqref{eq:ode_f(t,x)}, then as $ t\to \infty $ all trajectories $ x(t,  x_o) $ converge to $ x^\ast $ 
uniformly in $ t$ for all $ t_o \geq 0 $ and $ x_o $. 
A point $ x^\ast $ can be attractive  for \eqref{eq:ode_f(t,x)} without being an equilibrium of \eqref{eq:ode_f(t,x)} (we will use this fact extensively in the paper).

Given \eqref{eq:ode_f(t,x)}, denote $ g_s(t, \, x ) $ the translate of $ g(t, \, x)$: $ g_s(t, \, x ) = g(t+s, \, x )$. 
A (possibly time-dependent) system $ \dot x = \tilde g (t, \, x) $ is called a {\em limit system} of \eqref{eq:ode_f(t,x)} if there exists a sequence $ \{ s_k\} $, $ s_k \to \infty $ as $ k\to \infty$, such that $ g_{s_k}(t, \, x ) $ converges to $ \tilde g (t, \, x)$ \cite{Artstein1976}. 
An existence condition for a limit system $ \tilde g(t, \, x) $ is given in Lemma~1 of \cite{Lee2001}: when $ g(t, \, x ) $ is a uniformly continuous and bounded function, then there exists increasing and diverging sequences $ \{ s_k \} $ such that  on compact subsets of $ \mathbb{R}^n$ $ g_{s_k} (t, \, x ) $ converges uniformly to a continuous limit function $ \tilde g(t, \, x) $ on every compact of $ [0, \, \infty)$. 
In general the limit system may not be unique nor time-invariant. 
However, when it exists unique, then it must be autonomous \cite{Artstein1976,rouche2012stability} because all translates $ g_{s+s'} (t, \, x ) $ must have themselves a limit system hence the latter cannot depend on time.
The time-varying system \eqref{eq:ode_f(t,x)} is called {\em asymptotically autonomous} in this case.

The $ \omega$-limit set of $ x(t, \, x_o) $, denoted $ \Omega_{x_o} $, consists of all points $ x^\ast $ such that a sequence $ \{ t_k\}$, with $ t_k \to \infty $ when $ k\to\infty $, exists for which $ \lim_{k\to\infty} x(t_k, \, x_o) = x^\ast $. 
For time-varying systems, if a solution is bounded then the corresponding $ \Omega_{x_o} $ is nonempty, compact and approached by $ x(t,  x_o)$. 
However, it need not be invariant. 
Only for limit systems the invariance property may hold, although not necessarily (it may fail even for asymptotically autonomous systems, see \cite{Artstein1976}).

The following lemma is inspired by \cite{MuASJC}, Thm 2.1 and \cite{saberi1990global}, Prop.~5, and provides us with a suitable comparison function to be used later in the paper.
The proof of this Lemma and of all other results is in the Appendix.

\begin{lemma}
\label{lemma:comparison1}
Consider the scalar system 
\beq
\dot v = -\alpha (v) + \beta (v, t) + \zeta (t) , \qquad  v(t_o) = v_o \geq 0.
\label{eq:comparison1}
\eeq
If $ \alpha (v) \in \mathcal{K}_\infty^2 $, $  \beta \in \mathcal{KL}_\infty^{1,e} $ and $ \zeta \in \mathcal{L}^e$, then the solutions of \eqref{eq:comparison1} are all prolongable to $ \infty $  and bounded $ \forall \; v_o \geq 0$ and $\forall t_o \geq 0$. Furthermore,
\[
\lim_{t\to \infty } v(t) = 0 \qquad \forall \; v_o\geq 0, \quad \forall \; t_o \geq 0.
\]
\end{lemma}

It follows from  $ \alpha(t) \geq 0$, $ \beta(v, t) \geq 0 $, $ \zeta(t) \geq 0$ for all $ v>0 $ and $ \alpha (0) =0 $ that $ \mathbb{R}_+ $ is invariant for the system \eqref{eq:comparison1}. 
If we can show that \eqref{eq:comparison1} remains bounded for all times, then \eqref{eq:comparison1} is also forward complete for all $ v_o\geq 0$.
Express $ \alpha(v) \in \mathcal{K}_\infty^2 $ as $ \alpha(v) = a v^2 $, $ \beta(v, t) \in \mathcal{KL}_\infty^{1,e} $ as $ \beta(v, t) = b v e^{-\delta_1 t}   $ and $ \zeta \in \mathcal{L}^e$ as $ \zeta(t) = c e^{-\delta_2 t} $ for some $ a, \, b, \, c >0$. 
Informally, boundedness follows from the fact that the globally exponentially stable ``unperturbed'' system $ \dot v = - a v^2 $ has a higher order as $ v\to \infty $ than the ``perturbation'' $ b v e^{-\delta_1 t } + c e^{-\delta_2 t }$. 
More in detail, in $ \mathbb{R}_+$, for $ v>1 $ it is $ v < v^2 $, hence we can write 
\[
b v e^{-\delta_1 t } + c e^{-\delta_2 t } < ( b  e^{-\delta_1 t } + c e^{-\delta_2 t }) v \qquad \forall \, v>1, \quad \forall \, t\geq t_o
\]
or 
\[
\dot v < ( -a v + b e^{-\delta_1 t } + c e^{-\delta_2 t }) v \qquad \forall \, v>1, \quad \forall \, t\geq t_o
\]
meaning that for $ v > \max \left( 1, \, \frac{b e^{-\delta_1 t_o } + c e^{-\delta_2 t_o } }{a} \right) $ it is $ \dot v <0$, $ \forall \, t \geq t_o $, i.e., the solution of \eqref{eq:comparison1} remains  bounded $ \forall \, v_o\geq 0 $ and $ \forall \, t_o \geq 0$.

Furthermore, $ \beta(v, t) $ and $\zeta(t) $ continuous, decreasing in $ t$ with $ \beta(v, t) {\to 0} $ and $ \zeta(t) \to 0 $ as $ t \to \infty$, imply also that for any $ v_o>0 $ there exists a $ t_1 \geq t_o $ such that $ \forall \; t > t_1 $ $ \dot v(t) <0$.
Together with $ \mathbb{R}_+$-invariance, this implies that $ \lim_{t\to\infty} v(t) = d \geq 0$. 
To show that it must be $ d=0 $, 
let us assume by contradiction that $ d>0$. Then 
\[
\lim_{t\to \infty} \dot v(t) = \lim_{t\to \infty} (-\alpha (v) + \beta(v, t) + \zeta(t) ) = - \alpha(d) <0,
\]
meaning that there exists a $ t_2 > t_1 $ and a $ k \in (0, \, 1 ) $ such that 
\[
\dot v (t) < - k \alpha(d) <0 \qquad \forall \, t\geq t_2 .
\]
Applying the mean value theorem, we then have that $ \exists $ $ \tau \in [t_2, \, t] $ such that 
\[
\frac{v(t) - v(t_2)}{t-t_2} = \dot v (\tau) <  - k \alpha(d) <0 \qquad \forall \, t\geq t_2 ,
\]
from which it follows
\[
v(t) < - k \alpha(d) (t-t_2) + v(t_2) <0 \qquad \forall \, t\geq t_2 ,
\]
which is a contradiction since $ v\geq 0$.
\qed


\section{Problem formulation}
\label{sec:prob-form}
Consider a distributed dynamical system on a graph with $n$ nodes:
\beq \dot x = f(x)  ,\qquad x(0) = x_o ,
\label{eq:model}
\eeq  
where $ x= \begin{bmatrix} x_1 & \ldots & x_n \end{bmatrix}^T \in \mathbb{R}^n $ is a state vector and $ f\, : \, \mathbb{R}^n \to \mathbb{R}^n $ is a Lipschitz continuous vector field. Standing assumptions in this paper are that \eqref{eq:model} possesses a unique solution continuable on $ [0, \, \infty ) $ for all $ x_o \in \mathbb{R}^n $ and that interactions occur between first neighbors, i.e., 
\beq
\dot x_i = f_i ( x_i, \, x_j, \, j \in \mathcal{N}_i ) , \qquad i=1, \ldots, n
\label{eq:model_i}
\eeq
with $ \mathcal{N}_i $ the in-neighborhood of node $ i$. 

We are interested in cases in which the system \eqref{eq:model_i} has a globally asymptotically stable equilibrium point, perhaps depending on the initial conditions, i.e., $ \lim_{t\to \infty } x(t) = x^\ast $ for all $ x_o$, or $ \lim_{t\to \infty} x(t) = x^\ast(x_o) $.

The {\em privacy preservation problem} consists in using a system like \eqref{eq:model} to perform the computation of $ x^\ast $ in a distributed manner, while avoiding to divulgate the initial condition $ x_o $ to the other nodes. 
Clearly this cannot be achieved directly on the system \eqref{eq:model} which is based on exchanging the values $ x_i $ between the nodes. 
It can however be achieved if we insert a mask on the value $ x(t) $ which preserves convergence to $ x^\ast$, at least asymptotically. 
The masks we propose in this paper have the form of time-varying output maps.

\subsection{Output masks}

Consider a continuously differentiable time-varying output map
\beq
\begin{split}
h \, : \, \mathbb{R}_+ \times \mathbb{R}^n \times \mathbb{R}^m & \to  \mathbb{R}^n \\
(t, \, x, \, \pi ) & \mapsto  y(t) = h(t, x(t) , \pi) 
\end{split}
\label{eq:output1}
\eeq
where $ y = \begin{bmatrix} y_1 & \ldots & y_n \end{bmatrix}^T \in \mathbb{R}^n $ is an output vector of the same size as $x$, and $ \pi\in \mathbb{R}^{m} $ is a vector of parameters splittable into $n$ subvectors (not necessarily of the same dimension), one for each node of the network: $ \pi =\{ \pi_1,  \ldots , \pi_n \} $.

{In the following we refer to $h(t, x(t) , \pi)$ as an {\em output mask} and to $ y $ as a {\em masked output}. 
The state $ x$ of the system is first masked into $y$ and then transmitted to the other agents.
The original system \eqref{eq:model} can therefore be modified into the following {\em masked system:}}
\beq
\begin{cases}
\dot x & = f(y) \\ 
y & = h(t, x, \pi) .
\end{cases}
\label{eq:model_xy}
\eeq

We assume in what follows that the vector field $ f(\cdot) $ and the output trajectories $ y(t)$ are publicly known, while the state $ x$ and the output mask $ h(t, x, \pi) $ (functional form plus values of the parameters $ \pi$) are private to each agent. 

Let us introduce more in detail the output masks to be used in this paper.

\begin{definition}
\label{def:local_mask}
A $ C^1 $ output map $h$ is said a {\em local mask} if it has components that are local, i.e.,
\benu
\item[P1:] $ h_i(t, x, \pi) =  h_i(t, x_i, \pi_i ) \qquad i =1, \ldots, n$.
\eenu
\end{definition}
The property of locality guarantees that the output map $ h_i $ can be independently decided by each node, and can therefore remain hidden to the other nodes.  
Consequently, the problem of privacy preserving as it is formulated here cannot be cast as an observability problem, as each $ h_i(\cdot ) $ is unknown to the other agents. 
{To make things more precise, we introduce the following definition.
Consider the system \eqref{eq:model_xy}. Denote $ y(t, x_o) $ the output trajectory of \eqref{eq:model_xy} from the initial state $ x_o$.}
\begin{definition}
{An initial condition $ x_o $ is said {\em indiscernible from the output} if knowledge of the output trajectory $ y(t, x_o) $, $ t \in [t_o, \, \infty)$, and of the vector field $ f(\cdot )$ is not enough to reconstruct $ x_o$ in \eqref{eq:model_xy}.
It is said {\em discernible} otherwise.}
\end{definition}

{
\begin{remark}
In order to have discernible initial states, the following three conditions must all be satisfied
\benu[label=(\roman*)]
\item\label{item1} The exact functional form of the output mask $ h( \cdot) $ must be known;
\item\label{item2} The parameters $ \pi $ must be identifiable given the trajectory $ y(t, x_o) $ and the vector field $ f(\cdot)$;
\item\label{item3} The system \eqref{eq:model_xy} must be observable.
\eenu
For output masks, failure to satisfy \ref{item1} and \ref{item2} (or even just \ref{item2}) is enough to guarantee indiscernibility.
\end{remark}
}

In order to confound an agent monitoring the communications, the output map needs also to avoid mapping neighborhoods of a point $ x^\ast $ of \eqref{eq:model} (typically an equilibrium point) into themselves.
\begin{definition}
A $ C^1 $ output map $ h $ is said {\em not to preserve neighborhoods} of a point $ x^\ast $ if for all small $ \epsilon >0$, $ \|x_o- x^\ast \| <\epsilon $ does not imply $ \|h (0, x_o, \pi)- x^\ast \| < \epsilon$.
\end{definition}

Armed with these notions, we can now give the main definition of the paper.
\begin{definition}
\label{def:privacy_mask}
A $ C^1 $ output map $h$ is said a {\em privacy mask} if it is a local mask and in addition
\benu
\item[P2:] \label{def:mask1}  $ h_i(0, x_i, \pi_i) \neq x_i $ $ \forall \; x_i \in \mathbb{R}^n$, $ i=1, \ldots, n$;
\item[P3:] \label{def:mask2}  $ h(t,x, \pi) $ guarantees indiscernibility of the initial conditions;
\item[P4:] \label{def:mask3}  $ h(t,x, \pi) $ does not preserve neighborhoods of any $ x \in \mathbb{R}^n$;
\item[P5:] \label{def:mask4} $  h_i(t, x_i, \pi_i) $ strictly increasing in $ x_i$ for each fixed $ t$ and $ \pi_i$, $ i=1, \ldots, n$.
\eenu
\end{definition}
Property P5 
resembles a definition of $ \mathcal{K}_\infty $ function, but it is in fact more general: $ x=0 $ is not a fixed point of $ h $ for any finite $ t$, and $ h$ need not be nonnegative in $ x$.
It follows from Property P5 and locality that $ h $ is a bijection in $ x $ for each fixed $ t$ and $ \pi$, although one that does not preserve the origin. 
In many cases, it will be necessary to impose that the privacy mask converges asymptotically to the true state, i.e., that the perturbation induced by the mask is vanishing.
\begin{definition}
\label{def:vanishing_privacy_mask}
The output map $h$ is said a {\em vanishing privacy mask} if it is a privacy mask and in addition 
\benu
 \item[P6:] \label{def:mask5}$ | h_i(t, x_i, \pi_i ) - x_i | $ is decreasing in $ t$  for each fixed $ x_i$ and $ \pi_i$, and $ \lim_{t\to \infty } h_i(t, x_i, \pi_i ) = x_i $, $ i=1, \ldots, n$. 
 \eenu
\end{definition}

\subsection{Examples of output masks}
The following are examples of output masks.
\paragraph{Constant mask} 
\[
h_i(t, x_i, \pi_i ) = c_i  x_i, \qquad c_i > 1  
\]
(i.e., $ \pi_i = \{ c_i \} $). This local mask is not a privacy mask as e.g. properties~P2 and~P4 of Definition~\ref{def:privacy_mask} do not hold.
\paragraph{Linear mask}
\[
h_i(t, x_i, \pi_i ) = (1 + \phi_i e^{-\sigma_i t } ) x_i , \qquad \phi_i >  0, \quad \sigma_i >0 
\]
(i.e., $ \pi_i = \{ \phi_i, \, \sigma_i \}$). This local mask is not a proper privacy mask since $ h_i(0, 0, \pi_i ) =0 $ i.e. the origin is not masked. Notice that all homogeneous maps have this problem (and they fail to escape neighborhoods of $ x_i $).
\paragraph{Additive mask}
\[
h_i(t, x_i, \pi_i ) = x_i + \gamma_i e^{-\delta_i t }, \qquad \delta_i >0 , \quad \gamma_i \neq 0 
\]
(i.e., $ \pi_i = \{ \delta_i, \, \gamma_i \}$). This map does not preserve neighborhoods of $ x_i$, but it may fail to be a privacy mask, at least when the structure of $ h_i$ is known to an external agent, as shown by the following argument. Assume $ y(t, x_o )$ is transmitted to the other nodes and is publicly available, and so is $ f(\cdot)$. Then from system \eqref{eq:model_xy}, we can write
\[
\dot y_i = \dot x_i - \delta_i \gamma_i e^{-\delta_i t } = f_i(y) - \delta_i \gamma_i e^{-\delta_i t } .
\]
An agent that monitors the trajectory $ y(t, x_o ) $ and $ f(y(t, x_o)) $ can estimate $ \dot y(t)$ and hence also $ \delta_i \gamma_i e^{-\delta_i t } $. 
By looking at the observed exponentially decaying curve, the exponent $ \delta_i$ can be estimated, and therefore, from the value of the curve at $ t=0$, also $\gamma_i $ can be estimated.
So indiscernibility of the initial conditions is not guaranteed.
Clearly, for an agent unaware of the structure of $ h_i(\cdot) $ this mask is a fully fledged privacy mask.
\paragraph{Affine mask}
\[
h_i(t, x_i, \pi_i ) =c_i (x_i + \gamma_i e^{-\delta_i t }), \quad c_i > 1, \quad \delta_i>0, \quad  \gamma_i\neq 0 
\]
(i.e., $ \pi_i = \{ c_i, \, \delta_i, \, \gamma_i \}$). This is a privacy mask even when the structure of $ h(\cdot) $ is known to an external agent. In fact, repeating the same argument as above, one gets
\[
\dot y_i  = c_i (f_i(y) - \delta_i \gamma_i e^{-\delta_i t } ) \; \;  \Longrightarrow \; \;  f_i(y) - c_i^{-1}  \dot y_i  =  \delta_i \gamma_i e^{-\delta_i t } .
\]
Now knowledge of $ f(y(t)) $ and $ \dot y (t) $ is not enough to reconstruct $ c_i$, $ \delta_i $ and $ \gamma_i$ univocally. 
All other properties of Definition~\ref{def:privacy_mask} are also satisfied, as it is easy to check.
For instance, if $ x^\ast $ is any point of $ \mathbb{R}^n $ (typically an equilibrium of \eqref{eq:model}) and $ \| x(0) - x^\ast \|<  \epsilon$, then, denoting $ C = {\rm diag} (c_1, \ldots, c_n ) $, $ \| y(0) - x^\ast \| = \| C x(0) + C \gamma - x^\ast \| \leq \| C x(0) - x^\ast \| + \| C \gamma \| $ which in general does not belong to an $ \epsilon$-neighborhood of $ x^\ast $. 
Since $ \lim_{t\to\infty} h_i (t, \, x_i, \, \pi_i ) = c_i x_i $, this is however not a vanishing privacy mask.
\paragraph{Vanishing affine mask}
\beq
\begin{split}
h_i(t, x_i, \pi_i ) = & (1 + \phi_i e^{-\sigma_i t } ) (x_i + \gamma_i e^{-\delta_i t }), \\
&  \phi_i > 0, \quad \sigma_i >0, \quad \delta_i>0, \quad  \gamma_i\neq 0 
\end{split}
\label{eq:van-aff-mask-scalar}
\eeq
(i.e., $ \pi_i = \{ \phi_i, \, \sigma_i , \, \delta_i, \, \gamma_i \}$). This privacy mask is also vanishing.
Notice that in vector form, assuming all nodes adopt it, the vanishing affine mask can be expressed as
\beq
h(t, \, x, \pi)  = (I + \Phi e^{-\Sigma t} ) (x + e^{-\Delta t } \gamma)
\label{eq:output-mask-affine-II}
\eeq
where $ \Phi = {\rm diag}( \phi_1, \ldots, \phi_n ) $, $ \Sigma= {\rm diag}( \sigma_1, \ldots, \sigma_n ) $, $ \Delta = {\rm diag}( \delta_1, \ldots, \delta_n ) $, and $ \gamma = \begin{bmatrix} \gamma_1 & \ldots &  \gamma_n \end{bmatrix}^T $.

\subsection{Dynamically private systems}

\begin{definition}
\label{def:priv-sys}
The system \eqref{eq:model_xy} is called a {\em dynamically private} version of  \eqref{eq:model} if 
\benu
\item $ h $ is a privacy mask;
\item $ \lim_{t\to \infty} y(t) = x(t) $;
\item For each $ i=1,\ldots,n$, the integral $ \int_0^\infty f_i(y) dt $ cannot be estimated by an agent $ j \neq i$.
\eenu
\end{definition}

\begin{proposition}
\label{prop:no-equil}
If \eqref{eq:model_xy} is a dynamically private version of \eqref{eq:model}, then it cannot have equilibrium points.
\end{proposition}
\proof
The right hand side of the dynamics in \eqref{eq:model_xy} is autonomous.
Assume there exists $ y^\ast $ such that $ f(y^\ast) =0$. 
Since, from P5, $ h(\cdot ) $ is invertible in $ x$ for each $ t$, by the implicit function theorem, there exists an $ x^\ast (t) $ such that $ y^\ast = h(t, x^\ast(t), \pi) $. 
If $ x^\ast(t) $ is time-varying, then it is not an equilibrium point for \eqref{eq:model_xy}. 
If instead $ x^\ast $ is time-invariant, then, from $ \lim_{t\to \infty} y(t) = x(t) $, it must be $ x^\ast = y^\ast$. As this must be valid $ \forall \, t $, P2 is violated, hence also this case cannot happen in a privacy mask.
\qed

Lack of equilibria means that in a dynamically private system we cannot talk about stability.
The second condition in Definition~\ref{def:priv-sys} suggests that as long as $ f(\cdot ) $ is autonomous, a privacy-preserving masked system is asymptotically autonomous with the unmasked system as limit system. 
This can be shown to be always true if the output mask is vanishing. 

\begin{proposition}
\label{prop:asymtp-autonomous}
{Assume the solution of the dynamically private system \eqref{eq:model_xy} exists unique in $ [0, \, \infty ) $ $ \forall \; x_o \in \mathbb{R}^n$.}
If $ h $ is a vanishing privacy mask, then the system \eqref{eq:model_xy} is asymptotically autonomous with limit system \eqref{eq:model}.
\end{proposition}

\proof
We need to show that $ f(h(t, x, \pi)) \to f(x) $ as $ t \to \infty $ uniformly on compacts of $ \mathbb{R}^n $ \cite{Artstein1976}.
From P6 and $ h \in C^1 $, there exists an increasing, diverging sequence $ \{ t_k \} $ for which $ h_i (t_k,  x_i, \pi_i ) \to x_i $ as $ t_k \to \infty $, i.e., pointwise convergence holds.
In particular, for any $ \epsilon >0$, 
%
%
%
%
from pointwise convergence, there exists a $ \nu_o (x_i) $ such that, for all $ \nu>\nu_o $, $ |  h_i (t_{\nu}, x_i, \pi_i) - x_i | <\epsilon/2 $.
Pick two indexes $ \nu_1 = \nu_1(x_i) $, $ \nu_2=\nu_2(x_i) $ such that $ \nu_m > \nu_o$, $m = 1, \, 2 $. 
Then $ |  h_i (t_{\nu_1}, x_i, \pi_i) -  h_i (t_{\nu_2}, x_i, \pi_i)  | \leq |  h_i (t_{\nu_1}, x_i, \pi_i) - x_i | +  |  h_i (t_{\nu_2}, x_i, \pi_i) - x_i | \leq\epsilon/2 + \epsilon/2$. 
Selecting $ \nu_s = \sup_{x_i\in \mathcal{X}_i } \left\{ \nu_m (x_i) , \; m=1, \, 2 \right\}  $, then the Cauchy condition for uniform convergence applies and we have for any  integer $ \mu $ 
\[
\begin{split}
&  |  h_i (t_{\nu_s}, x_i, \pi_i) - x_i | \\
& \quad = \lim_{\mu \to \infty}  |  h_i (t_{\nu_s}, x_i, \pi_i) -  h_i (t_{\nu_s+ \mu}, x_i, \pi_i)  | \leq \epsilon .
\end{split}
 \]
 Hence, for a certain subsequence $ \{ t_\nu\} $ of $ \{ t_k \}$ it is $ {\rm sup}_{x_i \in \mathcal{X}_i } \left| h_i (t_\nu,  x_i, \pi_i ) - x_i \right| \to 0 $ as $ k\to \infty $, meaning that for $ h_i $ convergence is uniform on compacts. 
Since $ f_i $ is Lipschitz continuous, it is uniformly continuous and bounded on compacts. Hence Lemma~1 of \cite{Lee2001} holds, and by a reasoning identical to the one above, if $ \mathcal{X} $ is a compact of $ \mathbb{R}^n $ we have:
 \[
 {\rm sup}_{x \in \mathcal{X} } \left| f_i ( h (t_\nu,  x, \pi )) - f_i(x) \right| \to 0 \qquad \text{as} \quad \nu \to \infty .
 \]
The argument holds independently for any component $ f_i $. 
Asymptotic time-independence and uniform convergence on compacts to $ f(x) $ follow consequently.
\qed

The second condition in Definition~\ref{def:priv-sys} (for brevity: $ \lim_{t\to\infty} y_i(t) = y^\ast_i = x^\ast_i$), however, imposes an extra constraint on the problem, constraint that can lead to another form of disclosure of $ x(0)$. In fact, when $\int_0^\infty f_i(y) dt $ is known to an agent $ j\neq i $, then agent $j$ can use an expression like $ x_i^\ast = x_i(0) + \int_0^\infty f_i(y) dt $ to asymptotically estimate $ x_i(0) $ as $ x_i(0) = y^\ast_i - \int_0^\infty f_i(y) dt $. The third condition in Definition~\ref{def:priv-sys} is meant to avoid this possibility, and it is fulfilled if we make the following assumption \cite{Rezazadeh2018Privacy}.
\begin{assumption}
\label{ass1} {\rm (No overlapping neighborhoods)} 
The system \eqref{eq:model_xy} is such that $ \{ \mathcal{N}_i\cup \{ i \} \} \nsubseteq \{ \mathcal{N}_j \cup \{ j \}\} $, $ \forall \; i,\, j =1, \ldots, n$, $ i \neq j$. 
\end{assumption}
In fact, if the in-neighborhood of a node $ i$ is contained in that of another node $ j$, all $ y_k$ signals reaching $i$ are also available to $j$, hence $ \int_0^\infty f_i(y_k , \, k \in \mathcal{N}_i , y_i ) dt $ can be computed by $j$, and hence also $ x_i(0)$. Assumption~\ref{ass1} guarantees that no node has complete information of what is going on at the other nodes. 
Obviously, an alternative to Assumption~\ref{ass1} is for instance to keep $ f_i(\cdot) $ private to agent $ i$.


\section{Privacy-preserving average consensus}
\label{sec:av-consensus}

In the average consensus problem, $ f(x) = -L x $, with $ L$ a weight-balanced Laplacian matrix: $ L \bfone = L^T \bfone =0 $, with $ \bfone =\begin{bmatrix} 1 & \ldots & 1 \end{bmatrix}^T \in \mathbb{R}^n$.
When $L$ is irreducible, the equilibrium point is $ x^\ast(x_o) = \bfone^T x_o/n $. The system has a continuum of equilibria, described by $ {\rm span}(\bfone) $, and each $ x^\ast (x_o) $ is globally asymptotically stable in $ {\rm span} (\bfone)^\perp$, see \cite{Olfati2003Consensus}.

\begin{theorem}
\label{thm:consensus-c}
Consider the system
\beq
\dot x = - L x , \qquad x(0) =x_o 
\label{eq:consensus1}
\eeq
where $ L $ is an irreducible, weight-balanced Laplacian matrix, and denote $ \eta = \bfone^T  x_o /n $ its average consensus value.
Then $ x^\ast = \eta \bfone $ is a global uniform attractor on $ {\rm span} (\bfone)^\perp$ for the masked system 
\beq
\begin{split}
\dot x & = - L y  \\
y & = h(t, \,  x, \, \pi ) = \left(I + \Phi e^{-\Sigma t } \right)  \left( x - e^{-\Delta t} \gamma \right) .
\end{split}
\label{eq:consensus_xy}
\eeq
Furthermore, if Assumption~\ref{ass1} holds, then \eqref{eq:consensus_xy} is a dynamically private version of \eqref{eq:consensus1}.
\end{theorem}

\proof
Notice first that the system \eqref{eq:consensus_xy} can be written as
\beq
\dot x = - L  \left(I + \Phi e^{-\Sigma t } \right)  \left( x + e^{-\Delta t} \gamma \right)  ,
\label{eq:consensus2}
\eeq
from which it is clear that the system \eqref{eq:consensus_xy} cannot have equilibrium points. 
It is also clear from \eqref{eq:consensus2} that $ \bfone^T \dot x =0 $ i.e., also \eqref{eq:consensus_xy} obeys to the conservation law $  \bfone^T  x(t) =  \bfone^T x_o = \eta \bfone $. 
As in the standard consensus problem \cite{Olfati2003Consensus}, we can therefore work on the $ n-1$ dimensional projection subspace $ {\rm span}(\bfone)^\perp $ and consider the time-varying Lyapunov function for the ``displacement vector'' $ x - \eta \bfone \in {\rm span}(\bfone)^\perp $: 
\[
V(t, x) = ( x - \eta \bfone )^T \left(I + \Phi e^{-\Sigma t } \right)  ( x - \eta \bfone ) .
\]
From now on we assume that all calculations are restricted to $ {\rm span}(\bfone)^\perp $.
The derivative of $ V$ along the solutions of \eqref{eq:consensus_xy} is 
\beq
\begin{split}
& \dot V (t, \, x ) =  \pde{V}{x} \dot x + \pde{V}{t} \\
 = &  -  2 ( x - \eta \bfone )^T   \left(I + \Phi e^{-\Sigma t } \right)  L  \left(I + \Phi e^{-\Sigma t } \right)  \left( x + e^{-\Delta t} \gamma \right)   \\
& - ( x - \eta \bfone )^T   \left( \Sigma \Phi e^{-\Sigma t } \right)  ( x - \eta \bfone ) \\
= &  - ( x - \eta \bfone )^T    \left(I + \Phi e^{-\Sigma t } \right)  (L + L^T )  \left(I + \Phi e^{-\Sigma t } \right) \left( x - \eta \bfone \right) \\
& -  \eta ( x - \eta \bfone )^T  \left(I + \Phi e^{-\Sigma t } \right)  ( L + L^T)  \left(I + \Phi e^{-\Sigma t } \right)  \bfone  \\
& -  ( x - \eta \bfone )^T  \left(I + \Phi e^{-\Sigma t } \right)  ( L + L^T)   \left(I + \Phi e^{-\Sigma t } \right)   e^{-\Delta t} \gamma  \\
& - ( x - \eta \bfone )^T   \left( \Sigma \Phi e^{-\Sigma t } \right)  ( x - \eta \bfone ) .
\end{split}
\label{eq:consensus2b}
\eeq
Since $ \phi_i >0$, it is $ 1 + \phi_i e^{-\sigma_i t } \geq 1 $ $ \forall \, t \geq 0$, and $ I + \Phi e^{-\Sigma t }  $ is a positive definite diagonal matrix, for the first term of \eqref{eq:consensus2b} we have
\[
\begin{split}
 & ( x - \eta \bfone )^T    \left(I + \Phi e^{-\Sigma t } \right)  (L + L^T )  \left(I + \Phi e^{-\Sigma t } \right) \left( x - \eta \bfone \right) \\
 & \geq 
  ( x - \eta \bfone )^T ( L + L^T) ( x - \eta \bfone ) \geq \alpha_1(\| x- \eta \bfone \| ) > 0 
 \end{split} 
 \]
 for some function $ \alpha_1 \in \mathcal{K}_\infty^2 $.
The second term of \eqref{eq:consensus2b} is linear in $ \| x- \eta \bfone \| $, and from $ L\bfone = L^T \bfone =0$, we have
 \[
\begin{split}
& - \eta ( x - \eta \bfone )^T  \left(I + \Phi e^{-\Sigma t } \right)  ( L + L^T)  \left(I + \Phi e^{-\Sigma t } \right)  \bfone \\
& = - \eta ( x - \eta \bfone )^T  \left(I + \Phi e^{-\Sigma t } \right)  ( L + L^T)   \Phi e^{-\Sigma t }   \bfone \\
& \leq \beta_1  (\| x- \eta \bfone \| , \, t) 
\end{split}
\]
for some function $ \beta_1 \in \mathcal{KL}_\infty^{1,e} $. Similarly, for the third term of \eqref{eq:consensus2b},
\[
\begin{split}
&  - ( x - \eta \bfone )^T  \left(I + \Phi e^{-\Sigma t } \right)  ( L + L^T)   \left(I + \Phi e^{-\Sigma t } \right)   e^{-\Delta t} \gamma  \\
& \quad  \leq  \beta_2  (\| x- \eta \bfone \| , \, t) 
\end{split}
\]
for some $ \beta_2 \in \mathcal{KL}_\infty^{1,e} $.
Finally, the fourth term of \eqref{eq:consensus2b} is
\[
( x - \eta \bfone )^T   \left( \Sigma \Phi e^{-\Sigma t } \right)  ( x - \eta \bfone ) = \alpha_2(\| x- \eta \bfone \| , \, t) 
\]
for some $ \alpha_2 \in \mathcal{KL}_\infty^{2,e} $, i.e., it is positive definite for all finite $t$, and vanishes as $ t\to\infty$. 
hence there exists a $ \alpha  \in \mathcal{K}_\infty^{2} $ such that 
\[
\alpha (v) \geq \alpha_1(v) + \alpha_2 (v, t) >0 \quad \forall \; v \in \mathbb{R}^+.
\]
Denote $
\beta  (\| x- \eta \bfone \| , t )   \in \mathcal{KL}_\infty^{1,e} $ a proper majorization of $ \beta_j (\| x- \eta \bfone \| , t ) $, $j=1, \, 2$.
Since, for all $ t$, $ V$ is quadratic, positive definite, radially unbounded and vanishing in $ x = \eta \bfone $, there exists two class $ \mathcal{K}_\infty^2$ functions $ \alpha_3 $ and $ \alpha_4 $ such that 
\beq
\alpha_3 (\| x- \eta \bfone \| ) \leq V(t, x) \leq \alpha_4 (\| x- \eta \bfone \| )  .
\label{eq:consensus3}
\eeq
Also in this case we can apply the comparison lemma, using \eqref{eq:comparison1} with initial condition $ v(0) = V(0, x_o ) $, where $ x_o $ such that $ \bfone^T x_o /n = \eta$.
From Lemma~\ref{lemma:comparison1}, 
it follows that it must be $ \lim_{t\to\infty} V(t, x(t)) =0 $  for all $ x_o $ such that $ \bfone^T x_o /n = \eta$, hence from \eqref{eq:consensus3} $ \lim_{t\to\infty} \alpha_3(\| x- \eta \bfone \| ) =0 $ or $ \lim_{t\to\infty} x(t) =\eta \bfone$ for all $ x_o$ such that $ \bfone^T x_o /n = \eta$.
Since $ h(t, x, \pi)  = \left(I + \Phi e^{-\Sigma t } \right)  \left( x + e^{-\Delta t} \gamma \right)  $ is a privacy mask and Assumption~\ref{ass1} holds, \eqref{eq:consensus_xy} is a dynamically private version of \eqref{eq:consensus1}.
\qed

\begin{corollary}
\label{cor:consensus-autonomous}
The masked system \eqref{eq:consensus_xy} is asymptotically autonomous with \eqref{eq:consensus1} as limit system. 
The $ \omega$-limit set of \eqref{eq:consensus_xy} is given by $ \{ ( \bfone^T x_o /n ) \bfone    \} $ for each $ x_o $. 
\end{corollary}
\proof Asymptotic autonomy of \eqref{eq:consensus_xy} is shown using an argument identical to that of the proof of Proposition~\ref{prop:asymtp-autonomous}.
Convergence to the limit system \eqref{eq:consensus1} follows consequently. 
From expression \eqref{eq:consensus2}, the $ \omega$-limit set follows from $ x^\ast(x_o) $ being a uniform attractor for each $ x_o $. 
\qed

\begin{remark}
Even if  \eqref{eq:consensus1} has $ x^\ast =\eta \bfone  $ as a globally asymptotically stable equilibrium point in $ {\rm span}(\bfone)^\perp$, the masked system  \eqref{eq:model_xy} does not have equilibria and, because of the extra inhomogeneous term in the right hand side of \eqref{eq:consensus2}, not even stability of $ \eta \bfone $ is guaranteed. 
Nevertheless, $ x^\ast =\eta \bfone $ remains a global attractor for all trajectories of the system  in $ {\rm span}(\bfone)^\perp$.
\end{remark}

\begin{remark}
Since the evolution of the masked system \eqref{eq:consensus1} is restricted to the $ n-1$ dimensional subspace $ {\rm span}(\bfone)^\perp$, our masked consensus problem (as any exact privacy preserving consensus scheme) make sense only when $ n>2$. When $ n=2 $, in fact, each of the two agents can deduce the initial condition of the other from the value of $ \eta $ and the knowledge of its own initial state. 
\end{remark}

\begin{example}
\label{ex:private-consensus}
In Fig.~\ref{fig:consensus1} a private consensus problem is run among $ n=100 $ agents. 
Both $ x(t) $ (private) and $ y(t) $ (public) converge to the same consensus value $ \eta = \bfone^T x(0)/n$, but the initial condition $ y(0) $ does not reflect $ x(0)$, not even when $ x_i(0) $ is already near $ \eta $ ($h(\cdot ) $ does not preserve neighborhoods, see panel (c) of Fig.~\ref{fig:consensus1}). 
Notice that $ \bfone^T x(t)/n $ is constant over $t$, while $ \bfone^T y(t)/n $ is not, which confirms that the output masks indeed act as confounding factors.
Notice further that a standard Lyapunov function used for consensus, like $ V_{mm}(t) = \max_i(x_i(t)) - \min_i (x_i(t)) $, does not work in our privacy-preserving scheme (see panel (d) of Fig.~\ref{fig:consensus1}), which reflects the fact that the system \eqref{eq:consensus_xy} is not asymptotically stable in $ {\rm span}(\bfone)^\perp $.
Violation of $ \dot V_{mm}\leq 0$ is however not systematic but depending on the initial conditions, see Fig.~\ref{fig:consensus2}.
The convergence speed of the time-dependent part can be manipulated by selecting the factors $ \sigma_i $ and $ \delta_i $ appropriately.

\begin{figure*}[htb]
\begin{center}
\subfigure[]{
\includegraphics[angle=0, trim=1cm 7.5cm 9cm 1cm, clip=true, width=6cm]{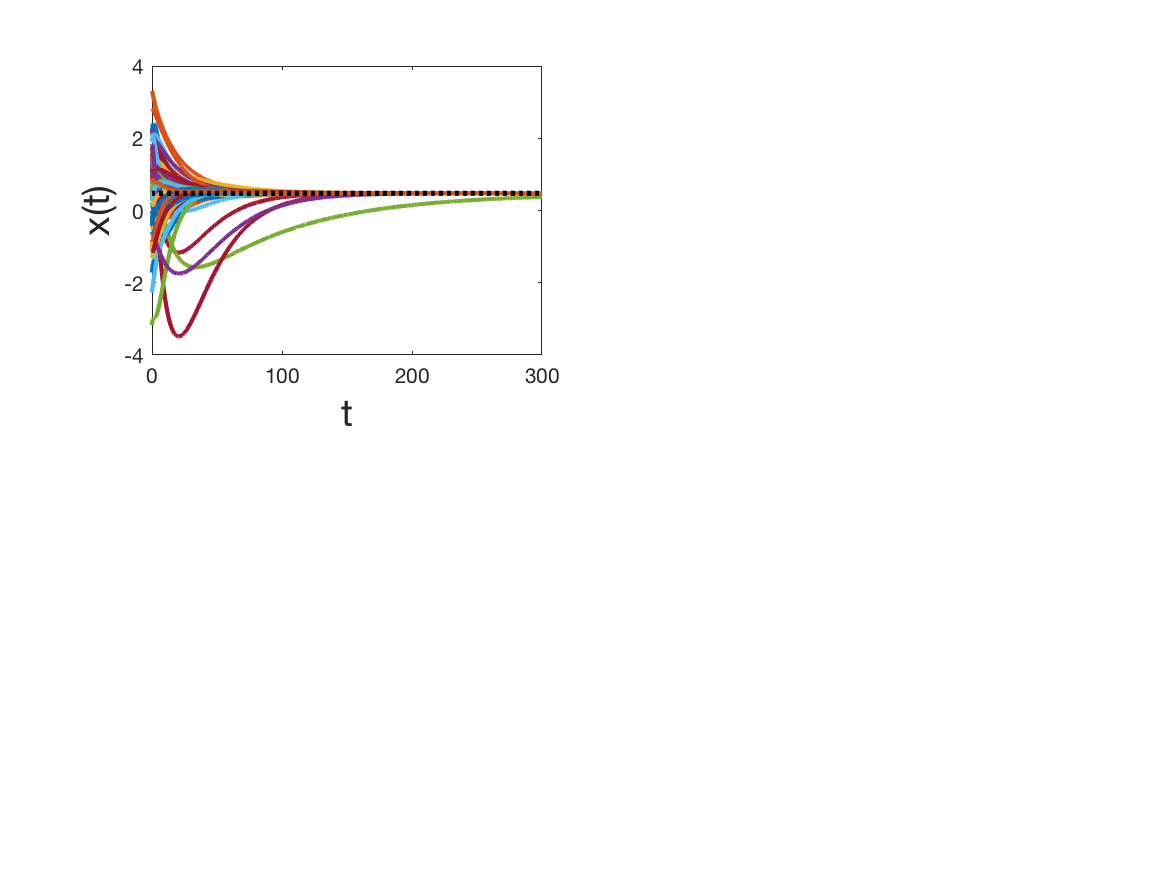}} 
\subfigure[]{
\includegraphics[angle=0, trim=1cm 7.5cm 9cm 1cm, clip=true, width=6cm]{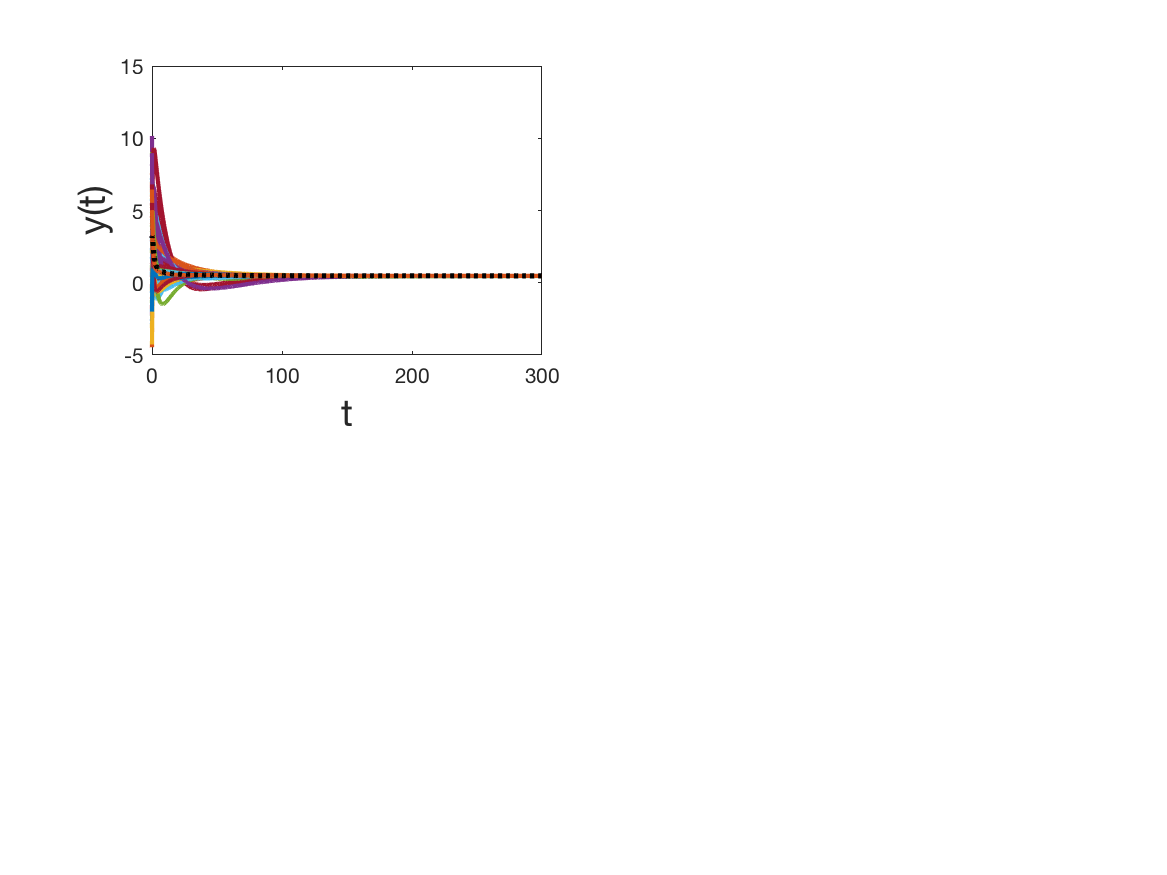}}\\
\subfigure[]{
\includegraphics[angle=0, trim=1cm 7.5cm 9cm 1cm, clip=true, width=6cm]{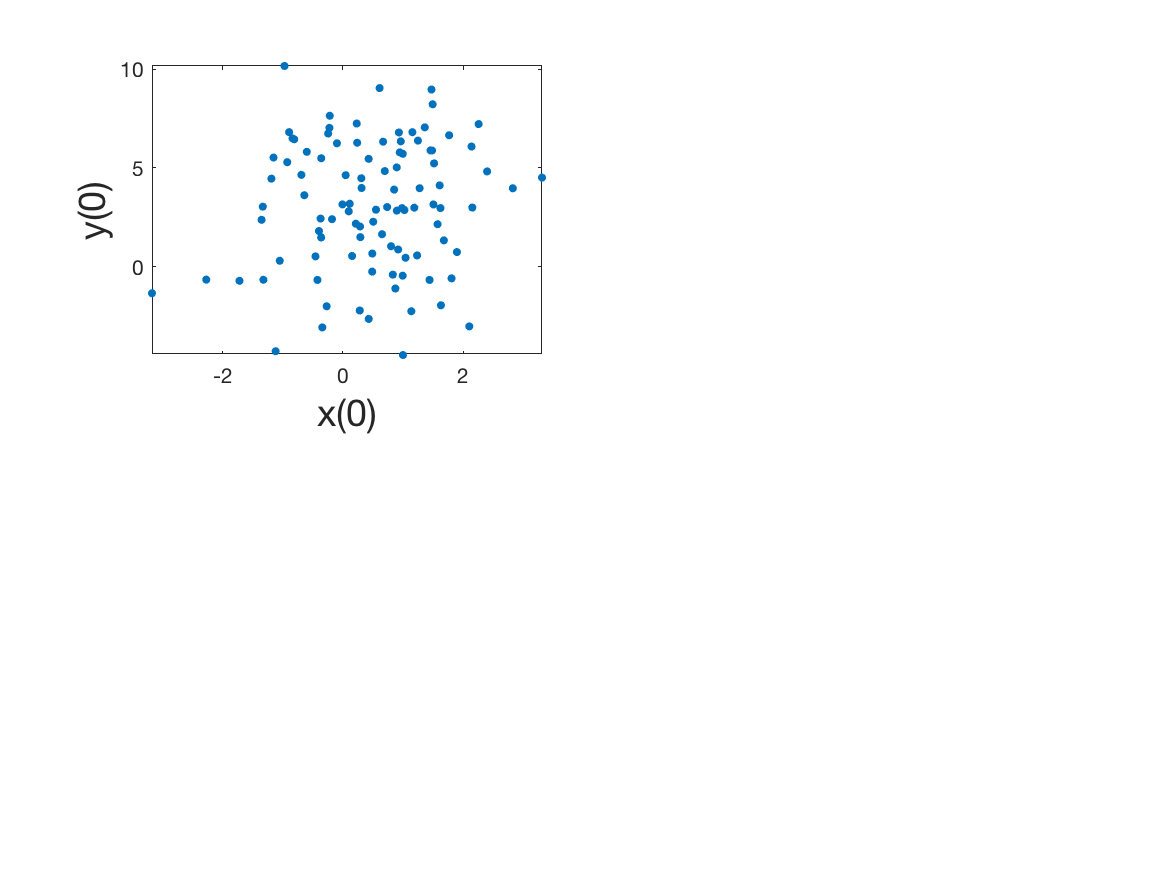}}
\subfigure[]{
\includegraphics[angle=0, trim=1cm 7.5cm 9cm 1cm, clip=true, width=6cm]{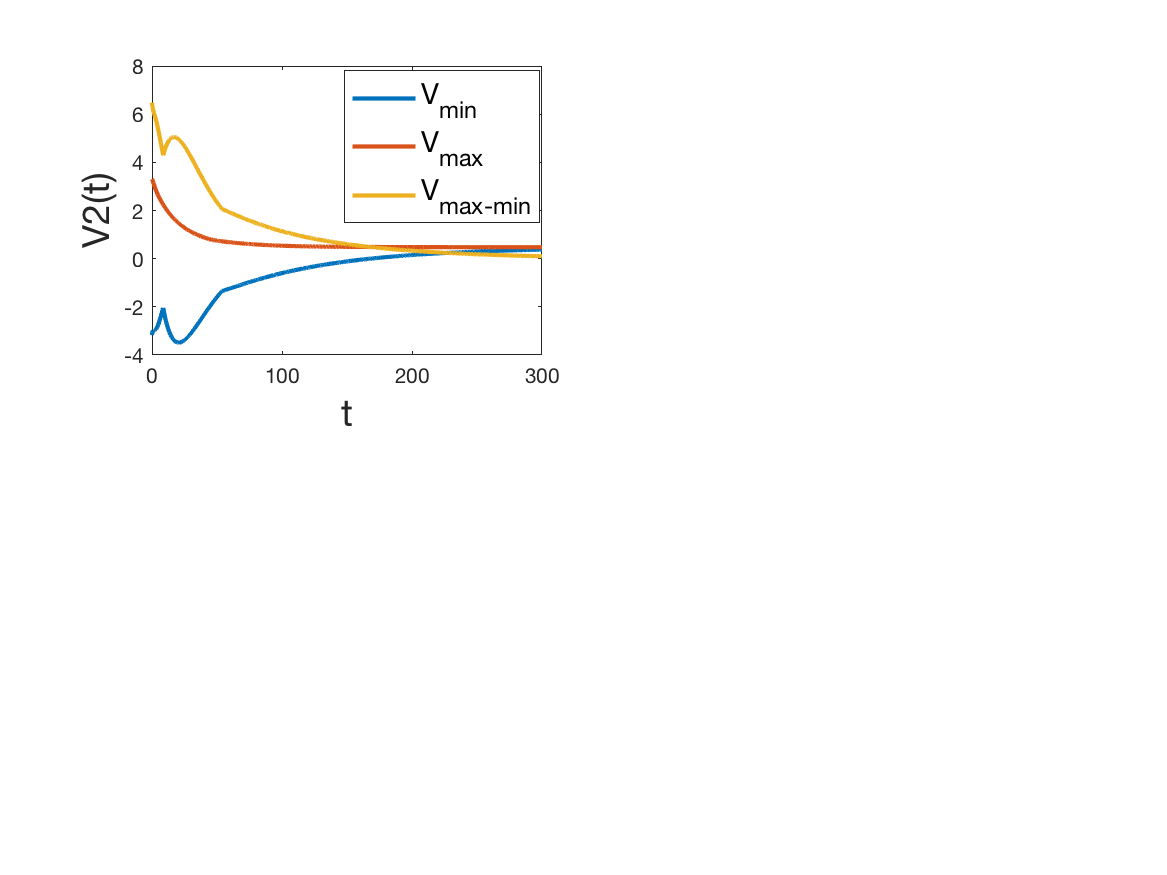}}

\caption{Privacy-preserving consensus of Example~\ref{ex:private-consensus}. (a): $ x(t)$; (b): $ y(t)$; (c): $ x(0) $ vs. $ y(0)$; (d): $ V_{mm}(t) = \max_i(x_i(t)) - \min_i (x_i(t)) $. The black dotted line in (a) resp. (b) represent $ \bfone^T x(t)/n $, resp. $ \bfone^T y(t)/n $.
}
\label{fig:consensus1}
\end{center}
\end{figure*}

\begin{figure*}[htb]
\begin{center}
\subfigure[]{
\includegraphics[angle=0, trim=1cm 7.5cm 9cm 1cm, clip=true, width=6cm]{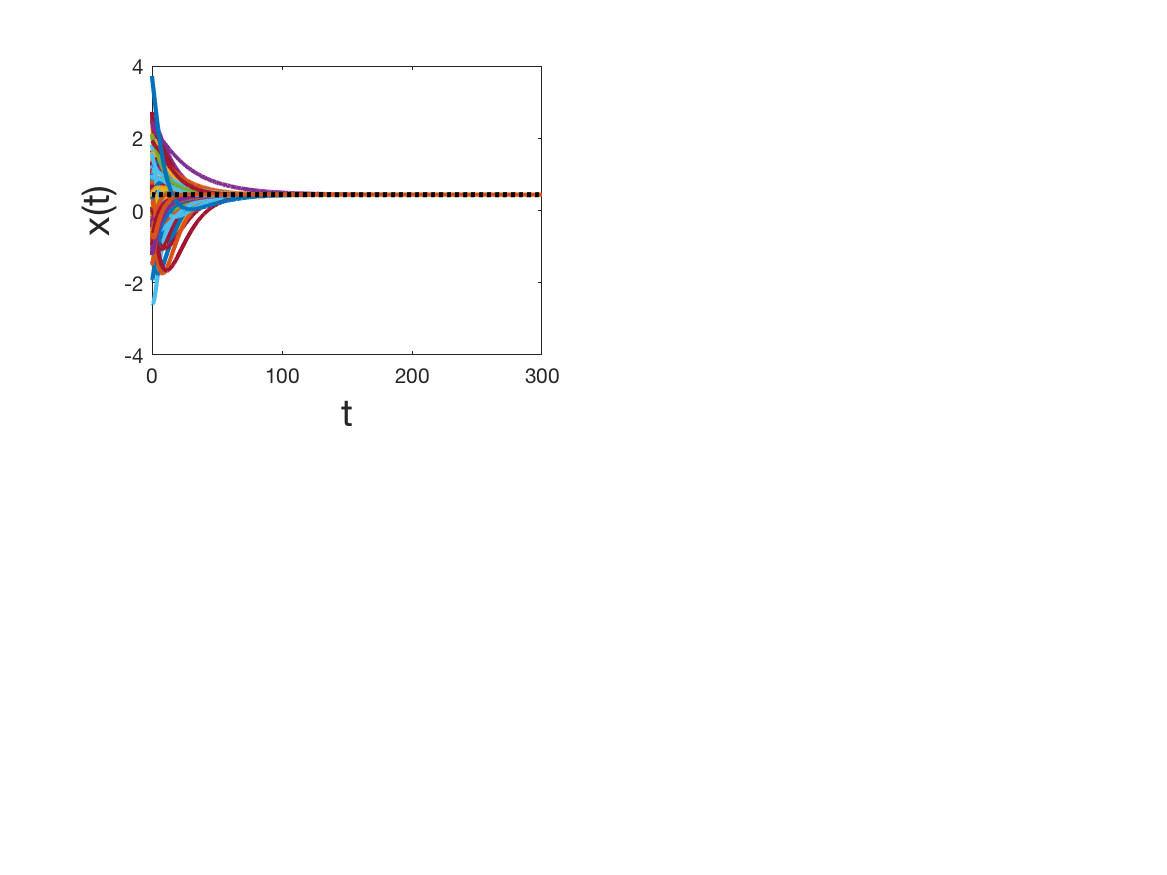}} 
\subfigure[]{
\includegraphics[angle=0, trim=1cm 7.5cm 9cm 1cm, clip=true, width=6cm]{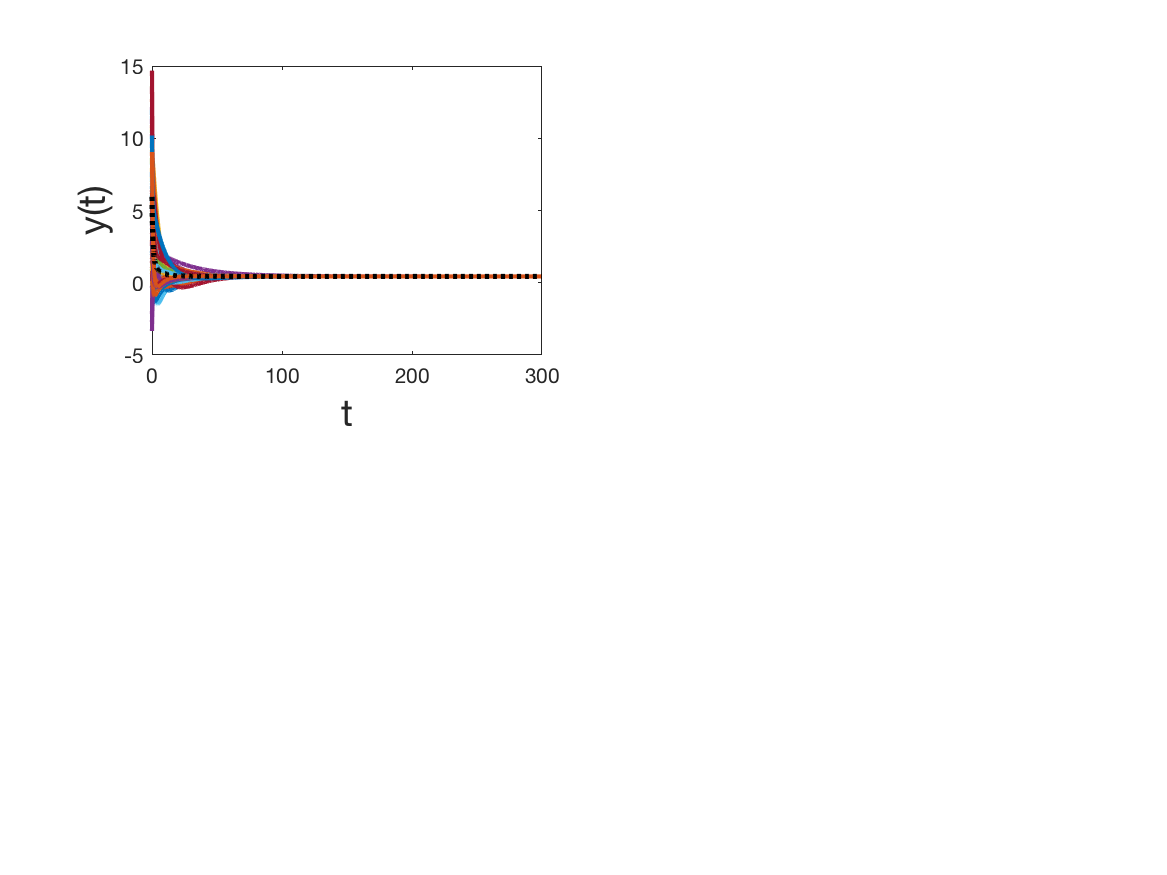}}\\
\subfigure[]{
\includegraphics[angle=0, trim=1cm 7.5cm 9cm 1cm, clip=true, width=6cm]{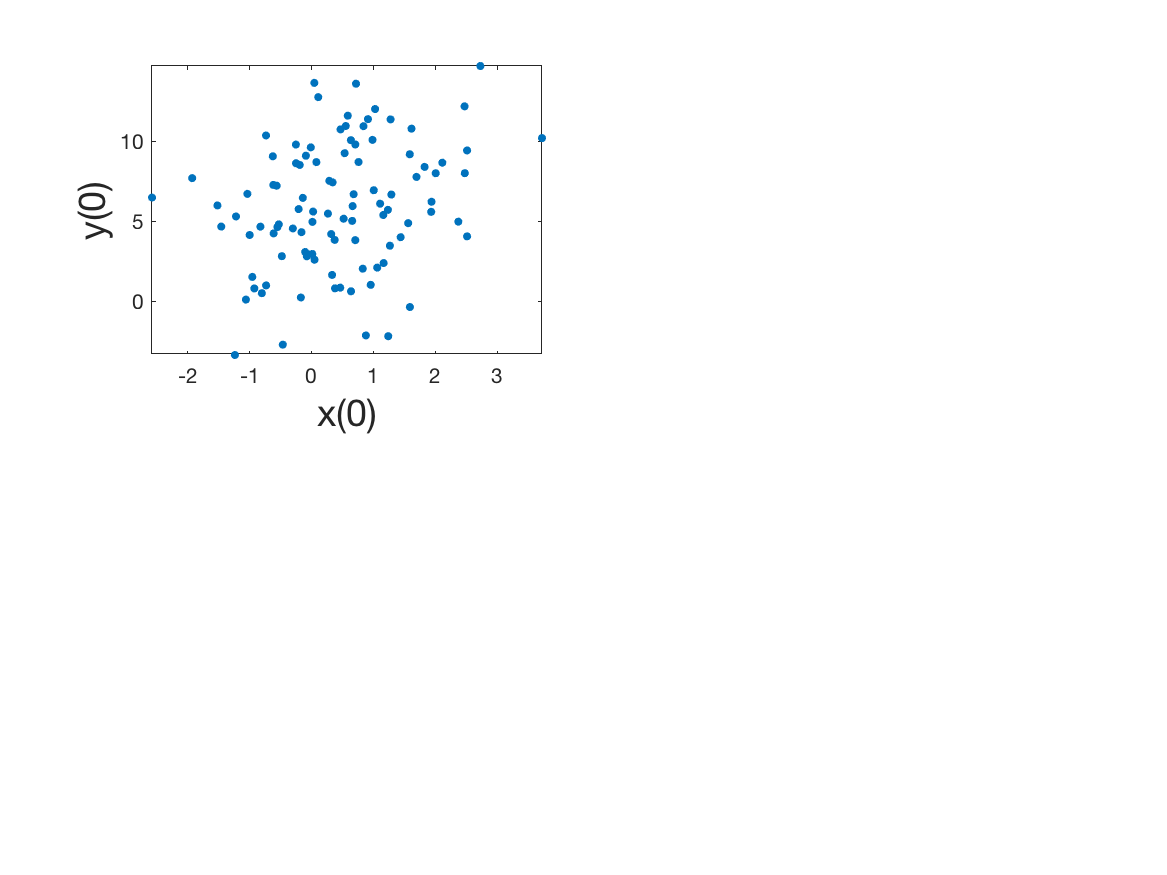}}
\subfigure[]{
\includegraphics[angle=0, trim=1cm 7.5cm 9cm 1cm, clip=true, width=6cm]{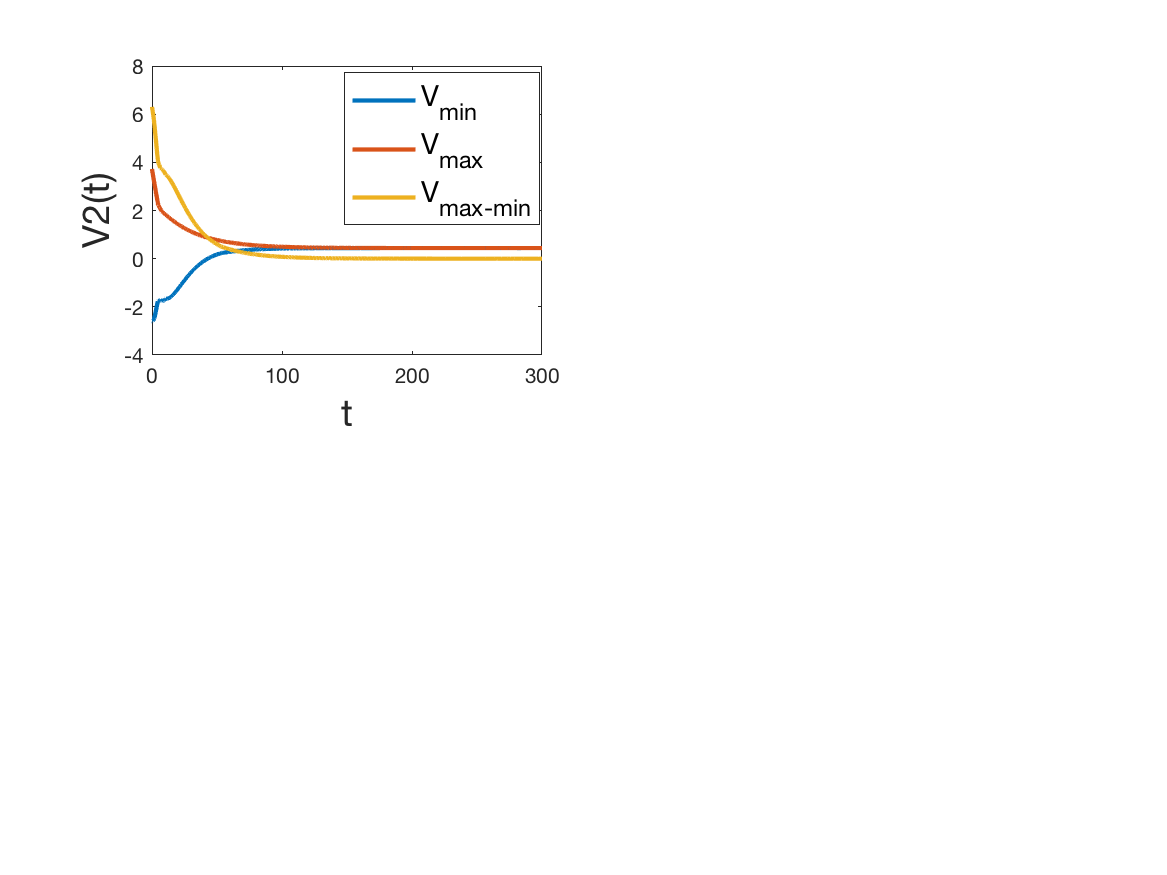}}

\caption{Another case of the privacy-preserving consensus of Example~\ref{ex:private-consensus}. (a): $ x(t)$; (b): $ y(t)$; (c): $ x(0) $ vs. $ y(0)$; (d): $ V_{mm}(t) = \max_i(x_i(t)) - \min_i (x_i(t)) $. The black dotted line in (a) resp. (b) represent $ \bfone^T x(t)/n $, resp. $ \bfone^T y(t)/n $.
}
\label{fig:consensus2}
\end{center}
\end{figure*}

\end{example}


\section{Conclusions}

When a distributed computation can be thought of as a multiagent dynamical system, then the problem of protecting the initial states of the agents can be formulated using classical tools from systems and control theory. 
For cases like average consensus, such system-theoretical framework provides a full solution: it exploits the naturally converging character of the original dynamics while at the same time hiding  the original initial states through output masks which render the system non-observable, and hence the true state non-estimable from the masked output. 
The framework, here applied only to average consensus for lack of space, is fairly general, and valid for a broad range of multiagent problems, see \cite{Altafini2018Privacy}. 
Notice how it is crucial for our method to deal with multiagent {\em dynamics}. 
Only with a dynamical system, in fact, can the extra layer introduced by the output mask decay and disappear over time, allowing convergence to the true state. 
In ``static'' contexts such as database query, where problems like privacy were originally formulated, our method is unlikely to provide any benefit. 


%
%
%

\begin{thebibliography}{10}

\bibitem{Alaeddini7963642}
A.~Alaeddini, K.~Morgansen, and M.~Mesbahi.
\newblock Adaptive communication networks with privacy guarantees.
\newblock In {\em 2017 American Control Conference (ACC)}, pages 4460--4465,
  May 2017.

\bibitem{Altafini2018Privacy}
C.~Altafini.
\newblock A system-theoretic framework for privacy preservation in multiagent
  dynamics.
\newblock {\em submitted for publication}, 2018.

\bibitem{Ambrosin:2017:OOB:3155100.3137573}
M.~Ambrosin, P.~Braca, M.~Conti, and R.~Lazzeretti.
\newblock Odin: Obfuscation-based privacy-preserving consensus algorithm for
  decentralized information fusion in smart device networks.
\newblock {\em ACM Trans. Internet Technol.}, 18(1):6:1--6:22, Oct. 2017.

\bibitem{Artstein1976}
Z.~Artstein.
\newblock Limiting equations and stability of nonautonomous ordinary
  differential equations.
\newblock In J.~LaSalle, editor, {\em The stability of dynamical systems}, CBMS
  Regional Conference Series in Applied Mathematics. SIAM, Philadelphia, 1976.

\bibitem{Cortes7798915}
J.~Cort\'es, G.~E. Dullerud, S.~Han, J.~L. Ny, S.~Mitra, and G.~J. Pappas.
\newblock Differential privacy in control and network systems.
\newblock In {\em IEEE 55th Conf. on Decision and Control}, pages 4252--4272,
  Dec 2016.

\bibitem{Dwork:2006:DP:2097282.2097284}
C.~Dwork.
\newblock Differential privacy.
\newblock In {\em Proceedings of the 33rd International Conference on Automata,
  Languages and Programming - Volume Part II}, ICALP'06, pages 1--12, Berlin,
  Heidelberg, 2006. Springer-Verlag.

\bibitem{Dwork:2014:AFD:2693052.2693053}
C.~Dwork and A.~Roth.
\newblock The algorithmic foundations of differential privacy.
\newblock {\em Found. Trends Theor. Comput. Sci.}, 9(3-4):211--407, Aug. 2014.

\bibitem{GUPTA20179515}
N.~Gupta, J.~Katz, and N.~Chopra.
\newblock Privacy in distributed average consensus.
\newblock {\em IFAC-PapersOnLine}, 50(1):9515 -- 9520, 2017.
\newblock 20th IFAC World Congress.

\bibitem{Huang:2012:DPI:2381966.2381978}
Z.~Huang, S.~Mitra, and G.~Dullerud.
\newblock Differentially private iterative synchronous consensus.
\newblock In {\em Proceedings of the 2012 ACM Workshop on Privacy in the
  Electronic Society}, WPES '12, pages 81--90, New York, NY, USA, 2012. ACM.

\bibitem{Lazzeretti6855039}
R.~Lazzeretti, S.~Horn, P.~Braca, and P.~Willett.
\newblock Secure multi-party consensus gossip algorithms.
\newblock In {\em 2014 IEEE International Conference on Acoustics, Speech and
  Signal Processing (ICASSP)}, pages 7406--7410, May 2014.

\bibitem{Lee2001}
T.-C. Lee, D.-C. Liaw, and B.-S. Chen.
\newblock A general invariance principle for nonlinear time-varying systems and
  its applications.
\newblock {\em IEEE Transactions on Automatic Control}, 46(12):1989--1993, Dec
  2001.

\bibitem{Liu2019Dynamical}
Y.~{Liu}, J.~{Wu}, I.~R. {Manchester}, and G.~{Shi}.
\newblock {Dynamical Privacy in Distributed Computing -- Part I: Privacy Loss
  and PPSC Mechanism}.
\newblock {\em arXiv e-prints}, page arXiv:1902.06966, Feb 2019.

\bibitem{Manitara6669251}
N.~E. Manitara and C.~N. Hadjicostis.
\newblock Privacy-preserving asymptotic average consensus.
\newblock In {\em 2013 European Control Conference (ECC)}, pages 760--765, July
  2013.

\bibitem{Markus1956}
L.~Markus.
\newblock Asymptotically autonomous differential systems.
\newblock In S.~Lefschetz, editor, {\em Contribution to the theory of nonlinear
  oscillations}, Annals of Mathematical Studies. Princeton Univ. Press,
  Princeton, 1956.

\bibitem{Mo7465717}
Y.~Mo and R.~M. Murray.
\newblock Privacy preserving average consensus.
\newblock {\em IEEE Transactions on Automatic Control}, 62(2):753--765, Feb
  2017.

\bibitem{MuASJC}
X.~Mu and D.~Cheng.
\newblock On the stability and stabilization of time-varying nonlinear control
  systems.
\newblock {\em Asian Journal of Control}, 7(3):244--255, 2005.

\bibitem{NOZARI2017221}
E.~Nozari, P.~Tallapragada, and J.~Cort\'es.
\newblock Differentially private average consensus: Obstructions, trade-offs,
  and optimal algorithm design.
\newblock {\em Automatica}, 81:221 -- 231, 2017.

\bibitem{Olfati2003Consensus}
R.~Olfati-Saber and R.~Murray.
\newblock Consensus problems in networks of agents with switching topology and
  time-delays.
\newblock {\em Automatic Control, IEEE Transactions on}, 49(9):1520 -- 1533,
  sept. 2004.

\bibitem{Pequito7039593}
S.~Pequito, S.~Kar, S.~Sundaram, and A.~P. Aguiar.
\newblock Design of communication networks for distributed computation with
  privacy guarantees.
\newblock In {\em 53rd IEEE Conference on Decision and Control}, pages
  1370--1376, Dec 2014.

\bibitem{Rezazadeh2018Privacy}
N.~Rezazadeh and S.~Kia.
\newblock Privacy preservation in a continuous-time static average consensus
  algorithm over directed graphs.
\newblock In {\em American Control Conference}, pages 5890--5895, 06 2018.

\bibitem{rouche2012stability}
N.~Rouche, P.~Habets, and M.~Laloy.
\newblock {\em Stability Theory by Liapunov's Direct Method}.
\newblock Applied Mathematical Sciences. Springer New York, 2012.

\bibitem{RuanW17}
M.~Ruan and Y.~Wang.
\newblock Secure and privacy-preserving average consensus.
\newblock {\em CoRR}, abs/1703.09364, 2017.

\bibitem{saberi1990global}
A.~Saberi, P.~Kokotovic, and H.~Sussmann.
\newblock Global stabilization of partially linear composite systems.
\newblock {\em SIAM Journal on Control and Optimization}, 28(6):1491--1503,
  1990.

\end{thebibliography}
%

\end{document}